\documentclass[12pt,a4paper]{article}
\usepackage{amsfonts,latexsym}
\usepackage{amsmath,amssymb}
\usepackage{graphicx,color}
\usepackage{subfigure}

\usepackage{multirow}
\usepackage[multiple]{footmisc}
\usepackage{booktabs}
\numberwithin{equation}{section}

\usepackage{cite}
\usepackage{bm}
\usepackage{dcolumn}
\renewcommand{\thefootnote}{\fnsymbol{footnote}}

\newcommand{\nn}{\nonumber}

\begin{document}
\vspace{12mm}

\begin{center}
{{{\Large {\bf f(R) black holes}}}}\\[10mm]

{Taeyoon Moon$^{a}$\footnote{e-mail address: dpproject@skku.edu},
Yun Soo Myung$^{b}$\footnote{e-mail address: ysmyung@inje.ac.kr},
and
Edwin J. Son$^{a}$\footnote{e-mail address: eddy@sogang.ac.kr}}\\[8mm]

{{${}^{a}$ Center for Quantum Space-time, Sogang University, Seoul, 121-742, Korea\\[0pt]
${}^{b}$ Institute of Basic Sciences and School of Computer Aided Science, Inje University Gimhae 621-749, Korea}\\[0pt]
}
\end{center}
\vspace{2mm}

\begin{abstract}
We study the $f(R)$-Maxwell black hole imposed by constant curvature
and its all thermodynamic quantities, which may lead  to the
Reissner-Nordstr\"om-AdS black hole by redefining   Newtonian
constant and charge.  Further, we obtain the $f(R)$-Yang-Mills black
hole imposed by constant curvature, which is related  to the
Einstein-Yang-Mills black hole in AdS space. Since there is no
analytic black hole solution in the presence of Yang-Mills field, we
obtain  asymptotic solutions. Then, we confirm the presence of these
solutions in a numerical way.
\end{abstract}
\vspace{5mm}

{\footnotesize ~~~~PACS numbers: 04.70.-s, 04.50.-h }


\vspace{1.5cm}

\hspace{11.5cm}{Typeset Using \LaTeX}
\newpage
\renewcommand{\thefootnote}{\arabic{footnote}}
\setcounter{footnote}{0}


\section{Introduction}
$f(R)$ gravities  as modified gravity theories
~\cite{NO,CLF,sf,NOuh} have much attentions as one of promising
candidates for explaining the current and future accelerating phases
in the evolution of  universe~\cite{SN}.  It is known that $f(R)$
gravities can be considered as general relativity (GR) with an
additional scalar field. Explicitly, it was shown that the
metric-$f(R)$ gravity is equivalent to the $\omega_{\rm BD}=0$
Brans-Dicke  theory with the potential, while the Palatini-$f(R)$
gravity is equivalent to the $\omega_{\rm BD}=-3/2$ Brans-Dicke
theory with the potential~\cite{FT}. Although the equivalence
principle test (EPT) in the solar system imposes a strong constraint
on $f(R)$ gravities, they may not be automatically  ruled out if the
Chameleon mechanism is employed to work.  It is shown that the EPT
allows $f(R)$ gravity models that are indistinguishable from the
${\rm \Lambda}$CDM model (GR with positive cosmological constant) in
the evolution of the universe~\cite{PS}. However, this does not
imply that there is no difference in the dynamics of
perturbations~\cite{CENOZ}.

On the other hand, the Schwarzschild-de Sitter black hole was
obtained for a positively constant curvature scalar in~\cite{CENOZ}
and other black hole solution  was recently found for a non-constant
curvature scalar~\cite{SZ}.  A black hole solution was obtained from
$f(R)$ gravities by requiring the negative constant curvature scalar
$R=R_0$~\cite{CDM}. If $1+f'(R_0)>0$, this black hole is similar to
the Schwarzschild-AdS (SAdS) black hole.  Also, its seems that there
is no sizable difference in thermodynamic quantities between $f(R)$
and SAdS black holes when using the Euclidean action approach and
replacing the Newtonian constant $G$ by $G_{\rm eff}=G/(1+f'(R_0))$.

In order to obtain the constant curvature black hole solution from
``$f(R)$ gravity coupled to the matter",  the trace of its
stress-energy tensor $T_{\mu\nu}$  should be zero.  Hence, two
candidates for the matter field are the Maxwell and Yang-Mills
fields. Concerning the $f(R)$-Maxwell black hole, the
authors~\cite{CDM} have made an mistake to show the correct
solution~\cite{cdme}.

In this work, we study the $f(R)$-Maxwell black hole and its all
thermodynamic quantities, which are similar to the
Reissner-Nordstr\"om-AdS (RNAdS) black hole when making appropriate
replacements. We obtain the topological $f(R)$-Maxwell black holes.
Importantly, we obtain the topological $f(R)$-Yang-Mills black
holes, which are similar to the topological Einstein-Yang-Mills
(dyonic) black holes in AdS space. Since there is no  analytic black
hole solution in the presence of Yang-Mills field, we obtain
asymptotic solutions. Then, we confirm the presence of these
solutions in a numerical way.

\section{$f(R)$-Maxwell black holes}
Let us first consider the action for $f(R)$ gravity with Maxwell
term in four dimensions
\begin{eqnarray} \label{Actionfm}
S_{fM}=\frac{1}{16\pi G}\int d^4 x\sqrt{-g} \Big[
R+f(R)-F_{\mu\nu}F^{\mu\nu} \Big].
\end{eqnarray}
From the variation of the above action (\ref{Actionfm}), the
Einstein equation of motion for the  metric can be written by
\begin{eqnarray} \label{equa}
R_{\mu\nu}\Big(1+f'(R)\Big)-\frac{1}{2}\Big(R+f(R)\Big)g_{\mu\nu}+
\Big(g_{\mu\nu}\nabla^2-\nabla_{\mu}\nabla_{\nu}\Big)f'(R)=2
T_{\mu\nu}
\end{eqnarray}
with the stress-energy tensor \begin{equation}
T_{\mu\nu}=F_{\mu\rho}F_{\nu}~^{\rho}-\frac{g_{\mu\nu}}{4}F_{\rho\sigma}F^{\rho\sigma}
~~~{\rm with}~T^{\mu}~_{\mu}=0.
\end{equation}
On the other hand, the Maxwell equation takes the form
\begin{equation}
\label{maxw}\nabla_{\mu}F^{\mu\nu}=0.
\end{equation}
Considering the constant curvature scalar $R=R_0$, the trace of
(\ref{equa}) leads to
\begin{eqnarray}
R_0\Big(1+f'(R_0)\Big)-2\Big(R_0 +f(R_0)\Big)=0\label{eqR}
\end{eqnarray}
which determines the negative constant curvature scalar  as
\begin{eqnarray}
R_0=\frac{2f(R_0)}{f'(R_0)-1}\equiv 4\Lambda_f<0. \label{eqCR}
\end{eqnarray}
Substituting this expression into (\ref{equa}) leads to the Ricci
tensor
\begin{equation} \label{riccit}
R_{\mu\nu}=\Lambda_f g_{\mu\nu}+\frac{2}{1+f'(R_0)}T_{\mu\nu},
\end{equation}
which implies that $R_{\mu\nu} \not=\Lambda_f g_{\mu\nu}$ (pure
AdS$_4$ space) unless $T_{\mu\nu}=0$.

We introduce a static spherically symmetric metric ansatz,
\begin{eqnarray}
ds^2=-N(r)dt^2+\frac{dr^2}{N(r)}+r^2 d\Omega^2_2
\end{eqnarray}
and a gauge field as a solution to (\ref{maxw}) \begin{equation}
A_{t}(r)=\frac{Q}{r_+}-\frac{Q}{r}\end{equation}
 which provides an electrically
charged black hole with $A_t(r_+)=0$.  Solving the Einstein equation
(\ref{equa}) together with the condition of constant curvature
scalar, we obtain the solution for a metric function
\begin{eqnarray} \label{metric0}
N(r)=1-\frac{2GM}{r}+\frac{Q^2}{(1+f'(R_0))r^2}-\frac{R_0}{12}r^2.
\end{eqnarray}
We note that the topological $f(R)$-Maxwell black hole solution is
also found to be
\begin{eqnarray}
N_k(r)=k-\frac{2GM}{r}+\frac{Q^2}{(1+f'(R_0))r^2}-\frac{R_0}{12}r^2
\end{eqnarray}
when considering the metric ansatz~\cite{BLP,Bir}
\begin{eqnarray}
ds_k^2=N_k(r)dt^2+N_k^{-1}(r)dr^2+r^2d\Sigma_k^2,\label{metric1}
\end{eqnarray}
with $d\Sigma_k^2 = d\theta^2+\sigma_k^{2}(\theta)d\varphi^2$. Here
$\sigma_{k}(\theta)$ denotes $\sin\theta,~\theta$ and $\sinh\theta$
for $k=1$ (spherical horizon), $~0$ (flat horizon), and $k=-1$
(hyperbolic horizon), respectively.

We could derive all thermodynamic quantities since the analytic
solution was known as (\ref{metric0}).  First of all,  the Hawking
temperature is calculated to be
\begin{eqnarray}
T_{H}(r_+,Q)=\frac{N'}{4\pi}|_{r \to r_+}= \frac{1}{4\pi}\Bigg[
\frac{1}{r_+}-\frac{Q^2}{(1+f'(R_0))r_+^3}-\frac{R_0r_+}{4}\Bigg].\label{metric1}
\end{eqnarray}
In order to compute other thermodynamic quantities, it would be
better to use  the Euclidean action approach~\cite{witten} because
we are working with $f(R)$ gravities. To make the action Euclidean,
the time coordinate should be made  imaginary by substituting
$t=i\tau$. In this case, to eliminate the conical singularity at the
horizon $r=r_+$, the coordinate $\tau$ should be periodic with the
period $\beta =1/T_H$. For this purpose, we have to calculate the
Euclidean action~\cite{MKP,DYK}
\begin{equation} \label{teact}
\triangle S^E_t=S^E_{fM}+S_{GH}+S_{ct}+S_{cF},
\end{equation}
where the Euclidean bulk action is
\begin{eqnarray}
 S^E_{fM}=-\frac{1}{16\pi G}\int d^4 x\sqrt{g_E}\Big[
R+f(R)-F_{\mu\nu}F^{\mu\nu} \Big].
\end{eqnarray}
Here $S_{GH}$ is the Gibbons-Hawking term to make the variation at
the boundary clear and $S_{ct}$ is the counter term for asymptotic
AdS$_4$ space. We are working with the canonical ensemble as the
fixed charge ensemble.  In this case, we need to introduce the
charge-fixing (cF) as a boundary surface term~\cite{MKP}
\begin{eqnarray}
S_{cF}=\frac{1}{4\pi G}\int d^3x\sqrt{h} F^{\mu\nu}n_{\nu}A_{\nu}
\end{eqnarray}
where $h_{ij}$ is the induced metric on the boundary surface and
$n_{\mu}$ is a radial unit vector pointing outwards. If one does not
introduce $S_{cF}$, one is working with the grand canonical
ensemble.  Also, the extremal black hole whose horizon is degenerate
is considered to be  the ground state in the canonical
ensemble~\cite{CEJM}. The location $r_+=r_e $ of extremal horizon is
determined by the condition of $T_H(r_e,Q)=0$. That is, in order to
derive the Helmholtz free energy, we must subtract the extremal mass
$M^f=M^f_e$ from (\ref{teact}). Taking into account all  leads to
\begin{eqnarray}
\triangle S^E_t-M^f_e&=&-\frac{
\beta(1+f'(R_0))}{48G}\Bigg[-12r_+-R_0r_{+}^3-\frac{36Q^2}{
(1+f'(R_0))r_+} \nn\\
&+&24\Big(r_e+\frac{Q^2}{
(1+f'(R_0))r_e}-\frac{R_0}{12}r_e^3\Big)\Bigg]\\
&\equiv& \beta F^f=\beta E^f-S^f_{BH}.
\end{eqnarray}
Here $F^f$ is the Helmhotz free energy, $\beta$ is the inverse of
the Hawking temperature, and $S^f_{BH}$ is the Bekenstein-Hawking
entropy. The energy, Bekenstein-Hawking entropy, and heat capacity
are given as
\begin{eqnarray}
E^f(r_+,Q)&=&\frac{\partial (\triangle S_E-M^f_e)}{\partial\beta}=M^f(r_+,Q)-M^f_e \nonumber \\
&=&\frac{(1+f'(R_0))r_+}{2G}\Bigg[1+\frac{Q^2}{(1+f'(R_0))r_+^2}-\frac{ R_0}{12} r_+^2\Bigg]-M^f_e,\\
S^f_{BH}&=&\beta E^f-\beta F^f=(1+f'(R_0))\frac{A (r_+)}{4G},\\
C^f(r_+,Q)&=&\Big(\frac{\partial E^f}{\partial T}\Big)_{Q} \nonumber \\
&=&\frac{2(1+f'(R_0))\pi r_{+}^2}{G}\Bigg[\frac{-4r_{+}^2
+\frac{4Q^2}{1+f'(R_0)}+R_0r_{+}^4}{4r_{+}^2
-\frac{12Q^2}{1+f'(R_0)}+R_0r_{+}^4}\Bigg],
\end{eqnarray}
where \begin{equation}
M^f_e=M^f(r_e,Q)=\frac{1+f'(R_0)}{3G}\Bigg[r_e+\frac{2Q^2}{r_e(1+f'(R_0))}\Bigg]
\end{equation}
 is the mass
of the extremal black hole and $A (r_+)=4\pi r_+^2$ is the horizon
area. In this case, $E$ measures the energy above the ground state.
\\
\begin{figure*}[t!]
   \centering
   \includegraphics{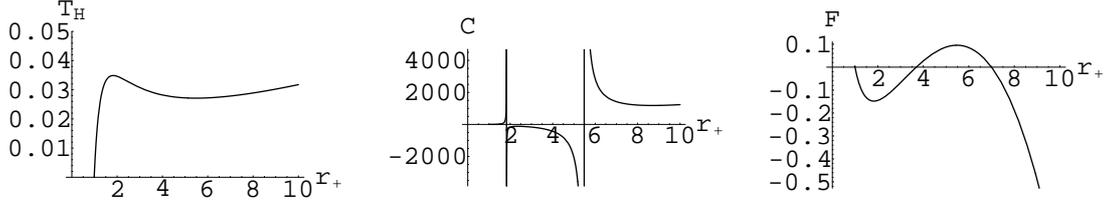}
\caption{Thermodynamic quantities of the $f(R)$-Maxwell black hole
as function of horizon radius $r_+$ with fixed $Q_f=1$, $\ell=10$,
and $G_{\rm eff}=1$: temperature $T_H$, heat capacity $C$, and
Helmhotz free energy $F$.} \label{fig4}
\end{figure*}
Considering  replacements of
\begin{equation}
\frac{G}{1+f'(R_0)}\to G_{\rm eff},~~R_0=4\Lambda_f \to
-\frac{12}{\ell^2},~~\frac{Q^2}{1+f'(R_0)} \to Q^2_f,
\end{equation}
the $f(R)$-Maxwell black hole becomes  the RNAdS black hole exactly.
In this case, the ADM mass $M^f$,  Hawking temperature $T_H$, and
the Bekenstein-Hawking entropy $S^f_{BH}$  take  compact forms
\begin{equation} \label{aas}
M^f(r_+,Q_f)=\frac{r_+}{2G_{\rm
eff}}\Bigg[1+\frac{Q_f^2}{r_+^2}+\frac{r_+^2}{\ell^2}\Bigg],~
T_{H}(r_+,Q_f)= \frac{1}{4\pi}\Bigg[
\frac{1}{r_+}-\frac{Q_f^2}{r_+^3}+\frac{3r_+}{\ell^2}\Bigg],~~S^f_{\rm
BH}=\frac{\pi r_+^2}{G_{\rm eff}}.
\end{equation}
Finally, the heat capacity $C^f$  and Helmholtz free energy $F^f$
are given by
\begin{eqnarray}\label{aac}
C^f(r_+,Q_f)&=& \frac{2\pi r_+^2}{G_{\rm
eff}} \Bigg[\frac{3r_+^4+\ell^2(r_+^2-Q_f^2)}{3r_+^4+\ell^2(-r_+^2+3Q_f^2)}\Bigg], \\
\label{aaf} F^f(r_+,Q_f)&=&\frac{1}{4G_{\rm
eff}r_+}\Bigg[r_+^2+3Q_f^2-\frac{r_+^4}{\ell^2}\Bigg]-M_e.
\end{eqnarray}
At this stage, we have to mention the other thermodynamic quantities
obtained directly from the metric function $N(r)$ in
(\ref{metric0}). In this case, all thermodynamic quantities of
$M,~S_{BH},~C,~F$ are obtained from the replacements as
\begin{equation}
R_0\to -\frac{12}{\ell^2},~~\frac{Q^2}{1+f'(R_0)} \to Q^2_f.
\end{equation}
There exists a slight difference in $M,~S_{BH},~C,~F$ between $G$ in
the direct method and $G_{\rm eff}$ in the Euclidean action
approach. The first law of thermodynamics is satisfied for both
cases
\begin{equation}
dM^f=T_{H}S^f_{BH},~~~dM=T_{H}S_{BH}.
\end{equation}
One curious quantity derived from $f(R)$ gravities is the
Bekenstein-Hawking entropy
\begin{equation}
S^f_{BH}=\Big[1+f'(R_0)\Big]\frac{\pi r_+^2}{G}
\end{equation}
which was also derived from the Wald method~\cite{FT}. On the other
hand, the conventional Bekenstein-Hawking entropy is
\begin{equation}
S_{BH}=\frac{\pi r_+^2}{G}.
\end{equation}
For example, if one uses $S^f_{BH}$ to check the first law of
thermodynamics, one immediately finds that it is not satisfied as
follows
\begin{equation}
dM \not=T_{H}S^f_{BH}.
\end{equation}
At this stage, there is no way to test  which approach provides  the
correct thermodynamic quantities for  $f(R)$-Maxwell black holes.
Anyway, the entropy issue should be resolved.

 The global features of thermodynamic quantities
are shown in Fig. 1 for $G_{\rm eff}=1=G$. Under this setting, there
is no difference between two approaches:
$M^f=M,~S_{BH}^f=S_{BH},~C^f=C,~F^f=F$. From the first and second
graphs, we observe the local minimum $T_H=T_0$ ($C$ blows up) at
$r_+=r_{0}$, in addition to the zero temperature $T_H=0$ ($C=0$) at
the extremal point of $r_+=r_e$ and the maximum value $T_H=T_m$ ($C$
blows up) at $r_+=r_{m}$ known as the Davies point. We note a
sequence of $r_e<r_m<r_0$. For $r_e<r_+<r_m$, the black hole is
locally stable because of $C>0$, while for $r_m<r_+<r_0$  it is
locally unstable ($C<0$). For $r_+>r_0$, the black hole becomes
stable because of $C>0$.
 Based on  the local stability, the $f(R)$-Maxwell  black holes are split
into small black hole (SBH) with $C>0$ being in the region of
$r_e<r_+<r_m$, intermediate black hole (IBH) with $C<0$ in the
region of $r_m<r_+<r_0$, and large black hole (LBH) with $C>0$ in
the region of $r_+>r_0$.

Importantly, the free energy from the last graph in Fig. 1 plays a
crucial role to test the phase transition. A black hole is globally
stable when $C>0$ and $F<0$. We note that $F=0$ at $r_+=r_e$,
because of $F=M-M_e-T_HS_{BH}$ with $T_H(r_e,Q_f)=0$. We observe two
extremal points for free energy: the local minimum $F=F_{min}$ at
$r_+=r_{m}$ and the maximum value $F=F_{max}$ at $r_+=r_{0}$. The
free energy is negative for $r_e<r_+<r_m$ and it increases  in the
region of $r_m<r_+<r_0$. For a point of $r_+=r_1>r_0$, it is zero
and remains negative for $r_+>r_1$.  The temperature of  $T=T_1$
(determined from the condition of $F=0$) at $r_+=r_1$ may play a
role of the critical temperature in Hawking-Page phase transitions.
The related phase transition was discussed in Ref.\cite{myungpt}.
It can be shown that the Hawking-Page phase transition II
 between SBH and LBH  unlikely occurs in the $f(R)$-Maxwell  black holes.

\section{$f(R)$-Yang-Mills black holes}

We consider the action of $f(R)$ gravity coupled to $SU(2)$
Yang-Mills field in four dimensions
\begin{eqnarray}
S_{fYM}=\frac{1}{16\pi G}\int d^4 x \sqrt{-g}\left\{
R+f(R)-F_{\mu\nu}^{a} F^{\mu\nu a}\right\},\label{Action1}
\end{eqnarray}
where
$F_{\mu\nu}^{a}=\partial_{\mu}A^{a}_{\nu}-\partial_{\nu}A^{a}_{\mu}+\epsilon^{abc}A_{\mu}^{b}
A_{\nu}^{c}$. From the  action (\ref{Action1}), the Einstein
equation of motion can be written by
\begin{eqnarray} \label{eqfym}
R_{\mu\nu}\Big(1+f'(R)\Big)-\frac{1}{2}\Big(R+f(R)\Big) g_{\mu\nu}+
\Big(g_{\mu\nu}\nabla^2-\nabla_{\mu}\nabla_{\nu}\Big)f'(R)=2T^{\rm
YM}_{\mu\nu}
\end{eqnarray}
with $T^{\rm YM}_{\mu\nu}$ the stress-energy tensor for the
Yang-Mills field.   For the constant curvature scalar $R=R_0$,
taking the trace of (\ref{eqfym})  leads to
\begin{eqnarray}
R_0\Big(1+f'(R_0)\Big)-2\Big(R_0 +f(R_0)\Big)=0\label{eqR}
\end{eqnarray}
which determines the constant curvature scalar as
\begin{eqnarray}
R_0=\frac{2f(R_0)}{f'(R_0)-1}\equiv 4\Lambda_f<0. \label{eqRC}
\end{eqnarray}
 Now we consider the topological
metric ansatz
\begin{eqnarray}
ds_k^2=-e^{2\phi(r)}N(r)dt^2+N^{-1}(r)dr^2+r^2d\Sigma_k^2,\label{metric1}
\end{eqnarray}
with  $d\Sigma_k^2 = d\theta^2+\sigma_k^{2}(\theta)d\varphi^2$. A
dyonic solution ansatz for Yang-Mills gauge field is given by
\begin{eqnarray}
A=\Big\{u(r)\tau_3
dt+\omega(r)\tau_{1}d\theta+\Big[\partial_{\theta}\sigma(\theta)\tau_3
+\sigma(\theta)\omega(r)\tau_2\Big]d\varphi\Big\},\label{gauge1}
\end{eqnarray}
where $u(r) [\omega(r)]$ describe the electric [magnetic] charged
configurations and $\tau_i$ is the Pauli spin matrices for $SU(2)$.

Substituting
 (\ref{metric1}) and (\ref{gauge1}) into the action
(\ref{Action1}), and  after variations with respect to
$N,~\phi,~\omega,~u$, one finds  their equations of motion:
\begin{eqnarray}
\delta_N
S;&&2r\Big(1+f'(R)\Big)\phi'-r^2\Big\{f'''(R)(R'(r))^2+f''(R)(-\phi'R'(r)+R''(r))\Big\}
\nn\\
&&\hspace*{14em}-4(\omega')^2-\frac{4e^{-2\phi}u^2\omega^2}{N^{2}}=0,\label{eqA}\\
\delta_{\phi}S;&&-2k+2rN'+2N-r^2f(R)+4N(\omega')^{2}+\frac{2(\omega^2-k)^2}{r^2}+2e^{-2\phi}
\Big(r^2(u')^2+\frac{2u^2\omega^2}{N}\Big)\nn\\
&&+2r^2Nf'''(R)(R'(r))^2+r
f''(R)\Big\{4NR'(r)+2rNR''(r)+rN'R'(r)\Big\}\nn\\
&& +r f'(R)\Big\{-2rN(\phi')^2-4N\phi'-r
N''-2rN\phi''-2N'-3rN'\phi'\Big\}=0,\label{eqphi} \\
 \delta_{\omega}S;
&&r^2N\omega''+r^2(N'+\phi'N)\omega'-\omega(\omega^{2}-k)
+\frac{e^{-2\phi}r^2u^2\omega}{N}=0,\label{eqomega}\\
\delta_{u}S;&&r^2u''+\left(-r^2\phi'+2r\right)u'-\frac{2\omega^2
u}{N}=0,\label{equ}
\end{eqnarray}
where the curvature scalar $R(r)$ is given by
\begin{equation}
R(r)=-\frac{1}{r^2}\Bigg[-2+rN'(4+3r\phi')+r^2N''+N(2+2r^2(\phi')^2+4r\phi'+2r^2\phi'')\Bigg].
\end{equation}
Note that the prime ($'$) in $f(R)$ and $N,\omega,\phi$ denotes the
differentiation with respect to $R$ and $r$, respectively. It is a
formidable task to solve the above four equations directly.
Therefore, we consider the constant curvature scalar which implies
that
\begin{equation}
\label{rrcon}R(r)=R_0,~~ R'(r)=R''(r)=0.
\end{equation}
Actually, we have  used the condition (\ref{rrcon}) to derive the
$f(R)$-Maxwell black holes in the previous section. Plugging
(\ref{rrcon}) into the four equations leads to  simplified equations
\begin{eqnarray}
\delta_N S;&&r\Big(1+f'(R_0)\Big)\phi'-2(\omega')^2-\frac{2e^{-2\phi}u^2\omega^2}{N^2}=0,\label{eqA}\\
\delta_{\phi}S;&&-2k+2rN'+2N-r^2f(R_0)+4N(\omega')^{2}+\frac{2(\omega^2-k)^2}{r^2}+2e^{-2\phi}
\left(r^2(u')^2+\frac{2u^2\omega^2}{N}\right)\nn\\
&&+r
f'(R_0)\Big(-2rN(\phi')^2-4N\phi'-r N''-2rN\phi''-2N'-3rN'\phi'\Big)=0,\label{eqphi}\\
\delta_{\omega}S;&&r^2N\omega''+r^2(N'+\phi'N)\omega'-\omega(\omega^{2}-k)
+\frac{e^{-2\phi}r^2u^2\omega}{N}=0\label{eqomega},\\
\delta_{u}S;&&r^2u''+\left(-r^2\phi'+2r\right)u'-2\frac{\omega^2
u}{N}=0.\label{equ}
\end{eqnarray}
Note that for $f(R_0)=f'(R_0)=0$, they reduce to those derived from
the topological Einstein-Yang-Mills theory~\cite{van}. It is well
known that there is no analytic black hole solution to the
Einstein-Yang-Mills theory. Hence, we can find either the asymptotic
solution with finite terms or the numerical solution.

First, we wish to derive the asymptotic solution at infinity of
$r\to \infty$. Equations (\ref{eqA})-(\ref{equ}) can be solved by
considering asymptotic forms for  metric and gauge field functions
up to $\frac{1}{r^5}$-order
\begin{eqnarray}
&&N(r)=k-\frac{2m(r)}{r}-\frac{R_0}{12}r^2,\label{A}\\
&&m(r)=M+\frac{M_1}{r}+\frac{M_2}{r^2}+\frac{M_3}{r^3}+\frac{M_4}{r^4}+\frac{M_5}{r^5}+O\left(\frac{1}{r^6}\right),\label{m:inf}\\
&&\omega(r)=\omega_{\infty}+\frac{\omega_1}{r}+\frac{\omega_2}{r^2}+\frac{\omega_3}{r^3}+\frac{\omega_4}{r^4}+
\frac{\omega_5}{r^5}+O\left(\frac{1}{r^6}\right),\label{w:inf}\\
&&u(r)=u_{\infty}+\frac{u_1}{r}+\frac{u_2}{r^2}+\frac{u_3}{r^3}+\frac{u_4}{r^4}+
\frac{u_5}{r^5}+O\left(\frac{1}{r^6}\right),\label{u:inf}
\end{eqnarray}
where $M, \omega_{\infty},~\omega_{1},~u_{\infty},$ and $~u_1$ are
five constants evaluated at infinity and  other $M_i$, $\omega_i$,
and $u_i$ are expressed in terms of these constants and $1+f'(R_0)$
appeared in Appendix A.  Let us compare $f(R)$-Yang-Mills (fYM)
black holes with Einstein-Yang-Mills (EYM) black holes. We observe
the relations of coefficients  between two black holes
\begin{eqnarray}
\label{subs1}M^{\rm fYM}_i&=&\frac{M^{\rm EYM}_i}{1+f'(R_0)},~{\rm
for}
~i=1,2,3,4 \\
\label{subs2}\omega_i^{\rm fYM}&=&\omega^{\rm EYM}_i,~~u_i^{\rm
fYM}=u^{\rm EYM}_i,~~~{\rm for}~ i=2,3,4.
\end{eqnarray}
It seems that  there is no longer simple relations between two black
holes for $i\ge 5$. Hence, it is not easy to derive  any concrete
form for thermodynamic quantities of $f(R)$-Yang-Mills black holes.
Exceptionally, the form of Hawking temperature can be derived to be
\begin{equation}
T^{\rm fYM}_H=\frac{1}{4\pi
r_+}\Bigg[k-\frac{R_0}{4}r_+^2-2m'(r_+)\Bigg]
\end{equation}
because it will be determined by the variables defined at horizon.
 Using
(\ref{mdath}), it takes the form
\begin{equation} T^{\rm fYM}_H=\frac{1}{4\pi
r_+}\Bigg[k-\frac{R_0}{4}r_+^2-\frac{(k-\omega_+^2)^2 +
r_+^4u_0^2e^{-2\phi_+}}{r_+^2(1+f'(R_0))}\Bigg]. \end{equation} In
the case of purely magnetic charged black hole with $u_0=0$ and
$f'(R_0)=0$, it reduces to Eq.(28) in Ref.\cite{van}. Here
$u_0=u'(r_+)$ may be  considered as a counterpart of
$A'_t(r_+)=Q/r_+^2$ in the $f(R)$-Maxwell black holes.
 Furthermore,
one finds the metric function $N$ ($m(r)\simeq M+M_1/r$) up to
$\frac{1}{r^2}$-order and gauge field functions $\omega$ and $u$ up
to $\frac{1}{r}$-order
\begin{eqnarray} \label{boundary1}
N(r)&\simeq&
k-\frac{2M}{r}+\Bigg\{\frac{Q_M^2+Q^2-\frac{R_0J^2}{6}-24\frac{
u_\infty^2\omega^2_\infty}{R_0}}{(1+f'(R_0))}\Bigg\}\frac{1}{r^2}-\frac{R_0}{12}r^2,
\\
\omega&\simeq&\omega_\infty+\frac{J}{r},~~u(r)\simeq
u_\infty-\frac{Q}{r} \label{boundary2}
\end{eqnarray}
with the Yang-Mills magnetic charge
$Q_M=k-\omega_\infty^2$~\cite{van} and the Yang-Mills electric
charge $Q$. Here we reset $\omega_1=J$ and $u_1=-Q$ to make a
connection to holographic super-conducting models using the AdS/CFT
correspondence~\cite{Gub} and ~\cite{MRT} for higher dimensional
cases. Using the holographic interpretation with $\omega_\infty=0$,
$u_{\infty}$ is the chemical potential, $Q$ is the electric charge,
and $J$ is the component of the current $J_i$ on the boundary at
infinity which is connected with the spontaneously broken part of
the bulk gauge symmetry.

We note that (\ref{boundary1}) and (\ref{boundary2}) with $Q_M^2=0$
and  $u_\infty=Q/r_+$[imposed by $u(r_+)=0]$ reduce to those of the
topological $f(R)$-Maxwell black holes when turning off the magnetic
charge gauge potential  and setting $u_i=0(i\ge 2)$. Especially, the
constant $\omega_1=J$ corresponds to an order parameter describing
the deviation from the Abelian solution of $f(R)$-Maxwell black
holes.

Finally, it is also interesting to explore the other case of purely
magnetic charged  black holes obtained by choosing $k=1,~u(r)=0$.
Its asymptotic solution appeared in Appendix B.  In the case of
Einstein-Yang-Mills black holes, these black holes are stable
against  gravitational and sphaleronic perturbations  for
$\omega_+>1/\sqrt{3}=0.577$ for large $|\Lambda|$~\cite{Win}.
Actually, the stability condition corresponds to that   a gauge
field $\omega(r)$ has no zero. Hence, we conjecture that purely
magnetic charged black holes in the $f(R)$-Yang-Mills theory has  a
similar property because for constant curvature scalar and
$1+f'(R_0)>0$ (no ghost condition), the $f(R)$-modification to the
Einstein-Yang-Mills black hole will be minimized. We will check in
the next section that the zero of $\omega(r)$ appears only for
$\omega_+<1/\sqrt{3}=0.577$. Furthermore, there exist nodeless
solutions for $k=0,1$ Einstein-Yang-Mills black holes~\cite{van},
which means that these black holes are stable.

In the next section, we will find  numerical solutions for $k=1,0$
dyonic black holes  and $k=1$ purely magnetic charged black holes.

\section{Numerical results}
We numerically solve \eqref{eqA}--\eqref{equ} with boundary
conditions of (\ref{boundary1}) and (\ref{boundary2})   using a
standard shooting method in {\it Mathematica}$^{\circledR}$7.
 The $k=1$ Einstein-Yang-Mills black holes was discussed in
 Ref.\cite{Win,BH}, while $k=0,-1$ Einstein-Yang-Mills black holes
 was found numerically in Ref.\cite{van}.

 In order to obtain numerical solutions,
we transform equations~\eqref{eqA}--\eqref{equ} into
\begin{align}
\phi'(r) &= \frac{2\Big(\omega^2 u^2 e^{-2\phi} + (\omega')^2 N^2\Big)}{r(1+f'(R_0))N^2}, \label{eq:phi} \\
m' (r)&= \frac{2r^2\Big(\omega^2 u^2 e^{-2\phi} + (\omega')^2 N^2\Big) + N\Big((k-\omega^2)^2+r^4((ue^{-\phi})')^2\Big)}{2r^2(1+f'(R_0))N} \notag \\
  &\quad +\frac{2ruu'\Big(\omega^2 u^2 e^{-2\phi} + (\omega')^2 N^2\Big)}{(1+f'(R_0))^2N^2} +
  \frac{2u^2\Big(\omega^2 u^2 e^{-2\phi} + (\omega')^2 N^2\Big)^2}{(1+f'(R_0))^3N^4}, \label{eq:m} \\
\omega''(r) &= -\phi' \omega' - \frac{N'\omega'}{N} - \frac{\omega(k-\omega^2)}{r^2N} - \frac{\omega u^2 e^{-2\phi}}{N^2}, \label{eq:w} \\
u''(r) &= \phi' u' - \frac{2u'}{r} + \frac{2u\omega^2}{r^2N},
\label{eq:u}
\end{align}
where we used the relation~\eqref{eqRC} to include $1+f'(R_0)$ only
as $f(R)$-gravity effects. First, let us develop the solution forms
near the non-degenerate horizon at $r=r_+$. Because of $N(r_+)=0$,
we derive a relation of $u(r_+)\omega(r_+)=0$ from Eq.~\eqref{eq:u}.
Choosing $\omega(r_+)=0$,  it is easily shown that $\omega'(r_+) =
\omega''(r_+) = 0$, which implies that $\omega(r)=0$. This is not
the case.  So we choose $u(r_+)=0$ instead, and the solution near
the non-degenerate horizon can be expanded as
\begin{align}
\phi(r) &= \phi_+ + \phi'(r_+) (r-r_+) + O(r-r_+)^2, \label{phi:nh} \\
m(r) &= m_+ + m'(r_+) (r-r_+) + O(r-r_+)^2, \label{m:nh} \\
\omega(r) &= \omega_+ + \omega'(r_+) (r-r_+) + O(r-r_+)^2, \label{w:nh} \\
u(r) &= u_0 (r-r_+) + \frac{u''(r_+)}{2} (r-r_+)^2 + O(r-r_+)^3,
\label{w:nh}
\end{align}
where the coefficients are determined by equations
\begin{align}
m_+ &= m(r_+) = \frac{r_+}{2} \left( 1 - \frac{R_0}{12} r_+^2 \right), \\
\phi'(r_+) &= \frac{2\omega_+^2\Big((k-\omega_+^2)^2 + r_+^4u_0^2e^{-2\phi_+}\Big)}{r_+(N'(r_+))^2(1+f'(R_0))}, \\
\label{mdath} m'(r_+) &= \frac{(k-\omega_+^2)^2 + r_+^4u_0^2e^{-2\phi_+}}{2r_+^2(1+f'(R_0))}, \\
\omega'(r_+) &= - \frac{\omega_+(k-\omega_+^2)}{r_+^2N'(r_+)}, \\
u''(r_+) &= -\frac{2u_0}{r_+} \left( 1 -
\frac{r_+^2\omega_+^2(4k-R_0r_+^2)}{4(N'(r_+))^2} \right)
\end{align}
which satisfy at the horizon $r=r_+$.  Since the metric function $N$
is zero at the horizon $r=r_+$ and it should be  positive outside
the horizon, we have to choose  a condition of $N'(r_+)>0$. This
restricts the range of $\omega_+$ through the inequality
\begin{equation}
2m'(r_+) < k - \frac{R_0 r_+^2}{4}
\end{equation}
which  yields a positiveness of $\omega'(r_+)>0$ for
$\omega_+(\omega_+^2-k)>0$, while $m'(r_+)>0$ for  $1+f'(R_0)>0$.
Note that $\phi'(r_+)$, $m'(r_+)$, $\omega'(r_+)$, and $u''(r_+)$
depend on $r_+$, $\phi_+$, $\omega_+$, and $u_0$, which means that
they  are four independent parameters describing  the near horizon
geometry of ~\eqref{phi:nh}--\eqref{w:nh}.  On the other hand, there
are five independent parameters of  $M$, $\omega_\infty$,
$\omega_1$, $u_\infty$, and $u_1$  describing  asymptotic region of
~\eqref{m:inf}--\eqref{u:inf}.  Remembering that
\eqref{eq:phi}--\eqref{eq:u} are two first- and two second-order
differential  equations, we need  six initial parameters to solve
the equations numerically at each boundary of horizon and asymptotic
infinity.  However, considering $N(r_+)=0$, we choose $u(r_+)=0$ and
then, \eqref{eq:w} leads  to a first-order differential  equation.
This is why four independent parameters is enough  to specify the
near horizon geometry of $f(R)$-Yang-Mills black hole. In addition,
we have time-rescaling symmetry so that we can replace $\phi$ by $
\phi+\phi_0$ without loss of generality. This means that either
$\phi_+$ or $\phi(\infty)$ can be cast to zero. Here  we choose
$\phi(\infty)=0$ to achieve an asymptotic AdS$_4$ space which is
similar to that of $f(R)$-Maxwell black hole solution. Hence,  the
asymptotic solution has five parameters less than  six of  general
analysis. In this manner, we show that $\phi_+$ is  not an arbitrary
parameter.

We are now  in a position to solve the initial value problem,  for
given $r_+$, $R_0$,  $\omega_+$, and $u_0$, by introducing a
specific form of $f(R)$ gravity as
 \begin{equation} \label{eqfr}
 f(R)=-\frac{\alpha c_1
\Big(\frac{R}{\alpha}\Big)^n}{1+\beta\Big(\frac{R}{\alpha}\Big)^n}
\end{equation} proposed in Ref.~\cite{HS} and setting $n=1$ for
simplicity in this work.  Imposing the constant curvature scalar
(\ref{eqRC}), Eq.(\ref{eqfr}) implies \begin{equation} c_1 =
\frac{\Big(1+\frac{\beta R_0}{\alpha}\Big)^2}{1+\frac{2\beta
R_0}{\alpha}}.
\end{equation}
The  Einstein limit exists in the constant curvature case, which
shows that $c_1\to2\beta\Lambda_f/\alpha$ as $\alpha\to0$.  Then,
$f(R)$ becomes a negative cosmological constant
($f(R)\to-2\Lambda_f$).
 In
what follows, however, we consider only a nonvanishing $\alpha$
case.
\begin{figure}[pbt]
\centering
  \subfigure[\ $k=+1$, $\omega_+=1.08$, and $u_0=0.33$
  ]{\label{fig:dyon:a}\includegraphics[width=0.45\textwidth]{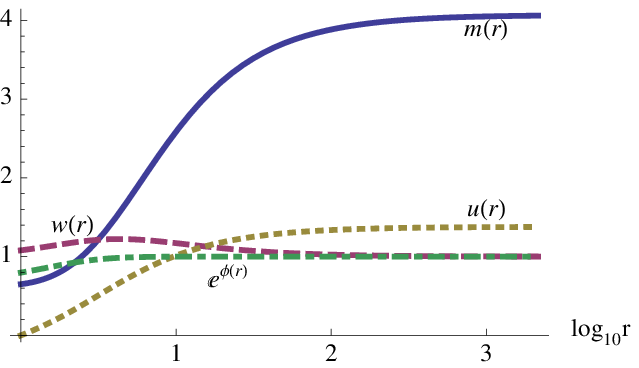}} \qquad
  \subfigure[\ $k=0$, $\omega_+=0.1$, and $u_0=0.77$
  ]{\label{fig:dyon:b}\includegraphics[width=0.45\textwidth]{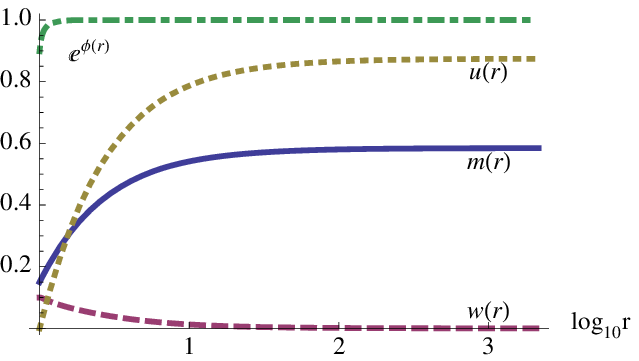}}
  \caption{\label{fig:dyon}
  The numerically solved functions $m$ (solid lines), $\omega$ (dashed lines), $u$ (dotted lines),
  and $e^\phi$ (dot-dashed lines) are depicted with respect to radius $\log_{10} r$  for
   $r_+=1$ and $R_0=-3.6$ ($\alpha = -0.9$, $\beta = 1.125$, $c_1 = 3.025$). \subref{fig:dyon:a}
   For $k=1$, we have $m_+=0.65$ and $\phi_+=-0.23$ with  $\phi(\infty)=0$. Then, we obtain $\omega_\infty=1.0$
   so that  $Q_M = (1-\omega_\infty^2)=0$. \subref{fig:dyon:b}
   For $k=0$, we get $m_+=0.15$ and $\phi_+=-0.11$ with $\phi(\infty)=0$.
    Then, we obtain $\omega_\infty=0$ so that  $Q_M=0$.
}
\end{figure}
\begin{figure}[pbt]
\centering
  \subfigure[\ $k=+1$, $\omega_+=1.08$, and $\phi_+=-0.03$]{\label{fig:mag:a}\includegraphics[width=0.45\textwidth]{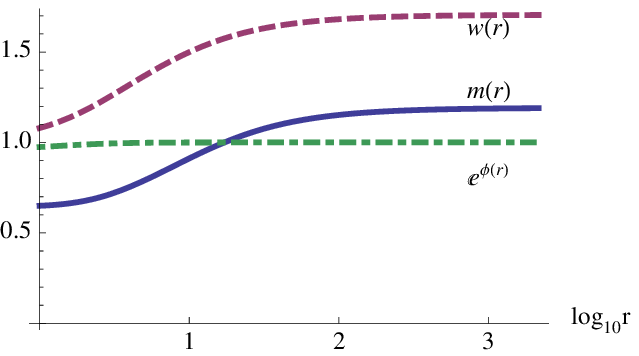}} \qquad
  \subfigure[\ $k=+1$, $\omega_+=0.1$, and $\phi_+=-0.01$]{\label{fig:mag:b}\includegraphics[width=0.45\textwidth]{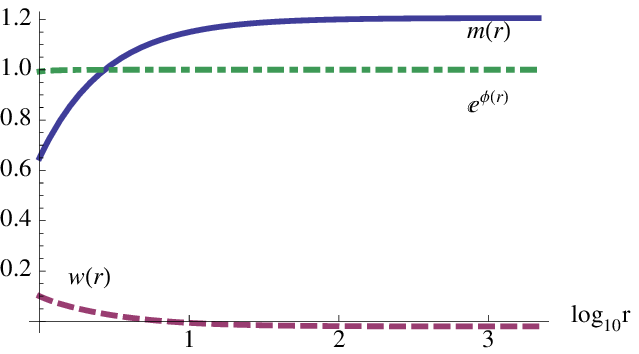}}
  \caption{\label{fig:mag}
  The numerically solved functions $m$ (solid lines), $\omega$ (dashed lines),
   and $e^\phi$ (dotdashed lines) of purely magnetic case ($u(r)=0$) are depicted with respect to radius
   $\log_{10}r$ for $r_+=1$, and $R_0=-3.6$ ($\alpha = -0.9$, $\beta = 1.125$, $c_1 = 3.025$).
    For convenience, $k=1$ case is plotted only, and we have $m_+=0.65$.
}
\end{figure}

The numerical solutions to \eqref{eq:phi}--\eqref{eq:u} are depicted
in Fig.~\ref{fig:dyon} graphically for $\alpha = -0.9$, $\beta =
1.125$, $c_1 = 3.025$, for which $R_0$ and $f'(R_0)$ are fixed to
either $R_0=-3.6$ and $f'(R_0)=-0.1$ or $R_0=1.8$ and $f'(R_0)=-10$.
In this work, since we are interested in asymptotically  AdS space
with $1+f'(R_0)>0$,  we choose the former of $R_0=-3.6$ and
$f'(R_0)=-0.1$.  Choosing the horizon radius to be  $r_+=1$, the
left and right figures are distinguished by specifying remaining
parameters $k$, $\omega_+$ and $u_0$: \subref{fig:dyon:a} $k=1$,
$\omega_+=1.08$ and $u_0=0.33$ \subref{fig:dyon:b} $k=0$,
$\omega_+=0.1$ and $u_0=0.77$. Furthermore, these numerical
solutions are being used to determine the form of  parameters in
asymptotic AdS space. Matching its asymptotic forms
(\ref{boundary1}) and (\ref{boundary2}),
 we find that for
\subref{fig:dyon:a}, $M \approx 4.07$, $\omega_\infty \approx 1.00$,
$\omega_1 \approx 2.63$, $u_\infty \approx 1.38$, and  $u_1 \approx
-4.14$,  while for  \subref{fig:dyon:b}, $M \approx 0.58$,
$\omega_\infty \approx 0.00$, $\omega_1 \approx 0.13$, $u_\infty
\approx 0.87$, and $u_1 \approx -0.87$. At this stage, we point out
that that the magnetic charge $Q_M=(k-\omega_\infty^2)$ vanishes for
both cases, so that their asymptotic geometry are similar to the
$f(R)$-Maxwell black holes.

It is also interesting to explore the other solution of purely
magnetic charged black holes  numerically by choosing $k=1,~u(r)=0$.
Considering its asymptotic solution appeared in Appendix B, we find
the numerical solutions.     For a given $f(R)$-form (\ref{eqfr}),
the numerical solutions to Eqs.(5.2)-(5.4) could be developed  for
the same values as in the dyonic black hole solution. Setting the
horizon at $r_+=1$, the two graphs in Fig. 3 are distinguished by
specifying a remaining parameter $\omega_+$: \subref{fig:mag:a}
$\omega_+=1.08$ \subref{fig:mag:b} $\omega_+=0.1$. Furthermore, this
numerical solutions are used to find the parameters in the
asymptotic solutions. We  find  $M \approx 1.19$, $\omega_\infty
\approx 1.17$, $\omega_1 \approx -2.54$ for \subref{fig:mag:a},
while $M \approx 1.21$, $\omega_\infty \approx -0.02$, $\omega_1
\approx 0.14$ for \subref{fig:mag:b}. The stability condition for
this black hole corresponds to the condition that a gauge field
$\omega(r)$ has no zero.  As is shown in the Fig. 3,  we  find that
the zero of $\omega(r)$ appears  for $\omega_+<1/\sqrt{3}=0.577$.
Hence, we conjecture that the stability condition  for the
Einstein-Yang-Mills black hole holds for the  $f(R)$-Yang-Mills
black holes with $1+f'(R_0)>0$.

\begin{table}[htdp]
\caption{Asymptotic forms of numerical solution for
$f(R)$-Yang-Mills black holes}
\begin{center}
\begin{tabular}{|r|c|c|c|c|}
  \hline
  \multirow{2}{*}{} & \multicolumn{2}{c|}{dyonic solution} & \multicolumn{2}{c|}{purely magnetic solution} \\
  \cline{2-5}
  & $k=+1$ & $k=0$ & $\omega_+=1.08$ & $\omega_+=0.1$ \\
  & ($\omega_+=1.08$, $u_0=0.33$) & ($\omega_+=0.1$, $u_0=0.77$) & ($k=+1$) & ($k=+1$) \\
  \hline
  $m(r)\approx$ & $4.07-\cfrac{18.87}{r}$ & $0.58-\cfrac{0.42}{r}$ & $1.19-\cfrac{4.18}{r}$ & $1.21-\cfrac{0.56}{r}$ \\
  $\omega(r)\approx$ & $1.00+\cfrac{2.63}{r}$ & $0.00+\cfrac{0.13}{r}$ & $1.71-\cfrac{2.54}{r}$ & $-0.02+\cfrac{0.14}{r}$ \\
  $u(r)\approx$ & $1.38-\cfrac{4.14}{r}$ & $0.87-\cfrac{0.87}{r}$ & 0 & 0 \\
  $\phi(r)\approx$ & $-\cfrac{15.57}{r^4}$ & $-\cfrac{0.01}{r^4}$ & $-\cfrac{3.58}{r^4}$ & $-\cfrac{0.01}{r^4}$ \\
  \hline
\end{tabular}
\end{center}
\label{tab:dyon}
\end{table}

\section{Discussions}
First of all, we summarize all numerical solutions to
$f(R)$-Yang-Mill black holes in Table 1. This table shows asymptotic
solution forms constructed in the numerical way: two dyonic
solutions for $k=1$ and $k=0$ black holes and two magnetically
charged black holes for $k=1$ and $\omega_+=1.08,~0.1$. The former
was developed to compare with the topological Einstein-Maxwell black
holes, while the latter was displayed to see the stability of
$f(R)$-Yang-Mills black holes.  We find that for $1+f'(R_0)>0$, the
$f(R)$-Yang-Mills black holes are similar to Einstein-Yang-Mills
black holes in AdS space. In this case, one may  develop the
second-order phase transition between $f(R)$-Maxwell and
$f(R)$-Yang-Mills black holes to explain the holographic
superconductor without Higgs field as in the Einstein
theory~\cite{Gub}.  The difference is that the AdS$_4$ space was
constructed  not by introducing a cosmological constant, but by
choosing an appropriate $f(R)$ function in (\ref{eqRC}).  Also,  it
seems that for $1+f'(R_0)>0$, the stability condition  of
magnetically charged Einstein-Yang-Mills black holes holds for
magnetically charged $f(R)$-Yang-Mills black holes.

The condition of $1+f'(R_0)>0$ is related to no ghost state for
graviton propagations on  AdS$_4$ space~\cite{NS,myungfR}, the
positiveness of effective Newton constant $G_{\rm eff}>0$ in
cosmological implications~\cite{FT,PS}, and a necessary condition
that $f(R)$ black hole becomes a type of  Schwarzschild-AdS black
hole~\cite{CDM}. In this work,  this condition is necessary to
obtain $f(R)$-Maxwell black hole and  to derive its thermodynamic
quantities.  Also, $f(R)$-Yang-Mills black holes requires this
condition to have asymptotic and numerical solutions. The other
condition of $f''(R_0)<0$ is not necessary to obtain the constant
curvature black hole solutions. However, this  condition may be
needed to be free from the Dolgov-Kawasaki instability related to
tachyonic mass~\cite{CLF,FT} in the perturbation analysis of
$f(R)$-Maxwell (Yang-Mills) black holes. We hope to  make a progress
on the perturbation analysis.

At this stage, we would like to mention the close connection between
$f(R)$ and Einstein black holes by rewriting the action
(\ref{Actionfm}) as
\begin{eqnarray}
\tilde{S}_{fM}=\int d^4 x\sqrt{-g} \left[\frac{1}{16\pi G}\left\{
R+f(R)\right\}-F_{\mu\nu}F^{\mu\nu} \right].
\end{eqnarray}
 In this case, Einstein equation  takes the form instead of
 (\ref{riccit})
\begin{eqnarray}
R_{\mu\nu}&=&\Lambda_f g_{\mu\nu}+\frac{8\pi
G}{1+f'(R_0)}\tilde{T}_{\mu\nu}
\end{eqnarray}
with $\tilde{T}_{\mu\nu}=4T_{\mu\nu}$.  Introducing a replacement of
$G_{eff} \to G/(1+f'(R_0))$, the above equation become
\begin{eqnarray}
R_{\mu\nu}&=&\Lambda_f g_{\mu\nu}+8\pi G_{eff}\tilde{T}_{\mu\nu}.
\end{eqnarray}
The solution is determined by
\begin{eqnarray}
\tilde{N}(r)&=&1-\frac{GM}{r}+\frac{16\pi
G_{eff}Q^2}{r^2}-\frac{R_0r^2}{12}
\end{eqnarray}
which is the same form as (\ref{metric0}) in the unit $16\pi G=1$.
Also, we may  make  such a replacement  for $f(R)$-Yang Mills theory
by rewriting (\ref{Action1}) as $\tilde{S}_{fYM}$. In this case, the
$M_i$, $\omega_i$, and $u_i$ in Appendix A including the
substitution rules (\ref{subs1}) and (\ref{subs2}) may be
conjectured by following $v=4\pi G/e^2$ with $e^2=1+f'(R_0)$ in
Ref.\cite{BH}, where $-F^2/4$ was used instead of $-F^2$.

Finally, we wish to comment on two points. One is to answer to the
question of ``is it possible to apply the reconstruction technique
of Ref.\cite{NOm} which was developed for pure $f(R)$ gravity to the
$f(R)$ with Maxwell (Yang-Mills) field?".  The answer is ``yes"
because the pure metric $f(R)$ gravity is equivalent to the
$\omega_{\rm BD}$ Brans-Dicke theory with the potential term.
Expressing $\phi=1+f'(R)$, we transform (\ref{Actionfm}) and
(\ref{Action1}) into the Brans-Dicke theory
$[\phi,V(\phi)=f-Rf'(R)]$ with Maxwell (Yang-Mills) field. As far as
the constant curvature scalar black hole solution is concerned, we
can obtain the same black hole solution from the reconstructed
Brans-Dicke theory with Maxwell (Yang-Mills) field~\cite{MMS1,MMS2}.
The other is to answer to the question of ``in the case of coupled
YM-$f(R)$ theory~\cite{BNO},
\begin{equation}
S_{YMf}=\frac{1}{16\pi G}\int d^4 x \sqrt{-g}\left\{ R-4\pi
G(1+f(R))F_{\mu\nu}^{a} F^{\mu\nu a}\Bigg[1+b\tilde{g}^2\ln
\Big[\frac{-0.5F_{\mu\nu}^{a} F^{\mu\nu
a}}{\mu^4}\Big]\Bigg]\right\}
\end{equation}
do we expect to obtain the similar solution?". In the case of
$f(R)=0$, the last term of the above action reduces to the effective
Lagrangian of $SU(N)$ Yang-Mills theory up to one-loop order with
\begin{equation}
b=\frac{1}{4}\frac{1}{8\pi^2}\frac{11}{3} N.
\end{equation}
Even though its equation of motions take complicated forms, we
expect to have similar numerical solution found here for the
constant curvature scalar black hole. This may be true because for
the constant curvature scalar black hole, the replacement of
$G_{eff} \to G/(1+f'(R_0))$ is expected to make the numerical
solution simple unless  $b$ log-term plays an important role.

Consequently, the $f(R)$-Maxwell (Yang-Mills) black holes imposed by
constant curvature scalar and $1+f'(R_0)>0$ are closely related to
the Einstein-Maxwell (Yang-Mills) black holes in AdS space.

{\bf Acknowledgments}

This work was supported by the National Research Foundation of Korea
(NRF) grant funded by the Korea government (MEST) through the Center
for Quantum Spacetime (CQUeST) of Sogang University with grant
number 2005-0049409.

\newpage

\section*{Appendix A: Coefficients for asymptotic solution to $f(R)$-Yang-Mills black holes}
\begin{eqnarray}
M_1&=&-\frac{k^2+u_1^2-\omega_1^2R_0
/6-2(k+12u_{\infty}^2/R_0)\omega_{\infty}^2+\omega_{\infty}^4}{2(1+f'(R_0))}
\nn\\
M_2&=&-\frac{2\omega_{\infty}(-12 u_1
u_{\infty}\omega_{\infty}/R_0+\omega_1(-k+\omega_{\infty}^2))}{1+f'(R_0)}
\nn\\
M_3&=&\frac{16 u_1 u_{\infty}\omega_1
\omega_{\infty}+R_0\omega_1^2(k-2\omega_{\infty}^2) +8
\omega_{\infty}^2(u_1^2+(k-\omega_{\infty}^2)^2-6u_{\infty}^2(k+2\omega_{\infty}^2)/R_0)}{R_0(1+f'(R_0))}
\nn\\
M_4&=&-\frac{1}{2(1+f'(R_0))}\left[\left(M-\frac{8 u_1
u_{\infty}}{R_0}\right)\omega_1^2+\frac{
48u_{\infty}\omega_{\infty}^2\left\{3Mu_{\infty}-4u_1(k-6u_{\infty}^2/R_0-2\omega_{\infty}^2)\right\}}{R_0^2}\right.
\nn\\
&&\left.+2\omega_1^3\omega_{\infty}-\frac{16\omega_1
\omega_{\infty}(2k^2+u_1^2-5k\omega_{\infty}^2+3\omega_{\infty}^4)}{R_0}+\frac{96\omega_1
\omega_{\infty}u_{\infty}^2(2k+\omega_{\infty}^2)}{R_0^2}\right]
\nn\\
M_5&=&\frac{1}{30(1+f'(R_0))}
\left[-6\omega_1^4+\frac{12\omega_1^2(21k^2+6u_1^2
-99k\omega_{\infty}^2+78\omega_{\infty}^4)}{R_0}
\right.\nn\\
&&\left.-\frac{576\omega_1^2u_{\infty}^2
(k-2\omega_{\infty}^2)}{R_0^2}
+\frac{288\omega_1\omega_{\infty}u_1u_{\infty}(9k-48
u_{\infty}^2/R_0-28\omega_{\infty}^2)}{R_0^2}
\right.\nn\\
&&\left.-\frac{360\omega_1\omega_{\infty}M(k-\omega_{\infty}^2))}{R_0}-
\frac{4320\omega_{\infty}^2Mu_1u_{\infty}+144\omega_{\infty}^2(13k-12\omega_{\infty}^2)(k-\omega_{\infty}^2)^2}
{R_0^2}
\right.\nn\\
&&\left.+\frac{144\omega_{\infty}^2u_1^2(13k-12\omega_{\infty}^2)}{R_0^2}-
\frac{17280\omega_{\infty}^2u_1^2u_{\infty}^2}{R_0^3}+\frac{41472\omega_{\infty}^2
u_{\infty}^4(2k+3\omega_{\infty}^2)}{R_0^4}
\right.\nn\\
&&\left.-\frac{1728\omega_{\infty}^2u_{\infty}^2(k-\omega_{\infty}^2)
(13k+6\omega_{\infty}^2)}{R_0^3}\right]-\frac{R_0}{60(1+f'(R_0))^2}\left[
\left(\omega_1^2+\frac{144u_{\infty}^2\omega_{\infty}^2}{R_0^2}\right)\times
\right.\nn\\
&&\left.\left(\omega_1^2-\frac{12(u_1^2+k^2-2k\omega_{\infty}^2+\omega_{\infty}^4)}{R_0}
-\frac{144\omega_{\infty}^2u_{\infty}^2}{R_0^2}\right)\right]
 \nn
\end{eqnarray}
\begin{eqnarray}
\omega_2&=&\frac{6}{R_0}\omega_{\infty}(k-12u_{\infty}^2/R_0-\omega_{\infty}^2)
\nn\\
\omega_3&=&\frac{6}{R_0}(-8u_1u_{\infty}\omega_{\infty}/R_0+\omega_1(k-4u_{\infty}^2/R_0-\omega_{\infty}^2))
\nn\\
\omega_4&=&-\frac{1}{2R_0}\left[4(3M+12u_1u_{\infty}/R_0)\omega_1
-\frac{12\omega_{\infty}(7k^2-2u_1^2+144u_{\infty}^4/R_0-10k\omega_{\infty}^2
+3\omega_{\infty}^4)}{R_0} \right.
\nn\\
&&\left. +6\omega_1^2\omega_{\infty}
+\frac{288\omega_{\infty}u_{\infty}^2(5k-4\omega_{\infty}^2)}{R_0^2}\right]
\nn\\
\omega_5&=&-\frac{1}{10R_0}\left[
6\omega_1^3+\frac{576\omega_{\infty}u_1u_{\infty}(11k-24u_{\infty}^2/R_0-7\omega_{\infty}^2)}{R_0^2}
+\frac{144\omega_{\infty}M(4k-60u_{\infty}^2/R_0-4\omega_{\infty}^2)}{R_0}
\right.\nn\\
&&\left.-\frac{12\omega_1\left\{39k^2-6u_1^2+144u_{\infty}^4/R_0^2-66k\omega_{\infty}^2+27\omega_{\infty}^4
-264u_{\infty}^2(k-2\omega_{\infty}^2)/R_0\right\}}{R_0}
\right]\nn\\
&&-\frac{\omega_1}{10(1+f'(R_0))}\left[-\frac{24u_1^2}{R_0}+3\omega_1^2-
\frac{24k^2-(48k+432u_{\infty}^2/R_0)\omega_{\infty}^2+24\omega_{\infty}^4}{R_0}\right]\nn
\end{eqnarray}

\begin{eqnarray}
u_2&=&-\frac{12}{R_0}u_{\infty}\omega_{\infty}^2
\nn\\
u_3&=&-\frac{4}{R_0}\omega_{\infty}(2u_{\infty}\omega_1+u_1\omega_{\infty})
\nn\\
u_4&=&-\frac{2}{R_0}\left[2u_1\omega_1\omega_{\infty}-\frac{144u_{\infty}^3\omega_{\infty}^2}{R_0^2}
+u_{\infty}\omega_1^2-\frac{24u_{\infty}\omega_{\infty}^2(-k+\omega_{\infty}^2)}{R_0}\right]
\nn\\
u_5&=&\frac{1}{5R_0}\left[48u_{\infty}\omega_{\infty}\left\{3M\omega_{\infty}+\omega_1(-6k+24u_{\infty}^2/R_0+
7\omega_{\infty}^2)\right\}\right]
\nn\\
&&+\frac{u_1}{30(1+f'(R_0))}\left[3\omega_1^2\left\{-1-\frac{12(1+f'(R_0))}{R_0}\right\}
\right.\nn\\
&&\left.+\frac{144\omega_{\infty}^2\left\{-3u_{\infty}^2+
(-6k+60u_{\infty}^2/R_0+4\omega_{\infty}^2)(1+f'(R_0))\right\}}{R_0^2}\right]
\nn
\end{eqnarray}
\newpage
\section*{Appendix B:  Asymptotic solution to a  magnetically charged black hole for
 $f(R)$-Yang-Mills theory}
In this case, SU(2) Yang-Mills gauge field is given by
\begin{eqnarray}
A=\left\{\omega(r)\tau_{1}d\theta+\left[\partial_{\theta}\sigma(\theta)\tau_3
+\sigma(\theta)\omega(r)\tau_2\right]d\varphi\right\}.\label{gauge}
\end{eqnarray}
Three equations of motion for $N$, $\phi$ and $\omega$ are given by
\begin{eqnarray}
\delta_N S;&&r(1+f'(R_0))\phi'-2(\omega')^2=0,\label{eqA1}\\
\delta_{\phi}
S;&&-2+2rN'+2N-r^2f(R_0)+4N(\omega')^{2}+\frac{2(\omega^2-1)^2}{r^2}\nn\\
&&\hspace*{3em}+r
f'(R_0)\Big(-2rN(\phi')^2-4N\phi'-r N''-2rN\phi''-2N'-3rN'\phi'\Big)=0,\label{eqphi1}\\
\delta_{\omega}S;&&r^2N\omega''+r^2(N'+\phi'N)\omega'-\omega(\omega^{2}-1)=0\label{eqomega1}.
\end{eqnarray}
The above equations  can be solved by assuming asymptotic forms up
to $\frac{1}{r^5}$-order
\begin{eqnarray}
&&N=1-\frac{2m}{r}-\frac{R_0}{12}r^2,\label{A}\\
&&m=M+\frac{M_1}{r}+\frac{M_2}{r^2}+\frac{M_3}{r^3}+\frac{M_4}{r^4}+\frac{M_5}{r^5}+O\left(\frac{1}{r^6}\right),\label{m:inf1}\\
&&\omega=\omega_{\infty}+\frac{\omega_1}{r}+\frac{\omega_2}{r^2}+\frac{\omega_3}{r^3}+\frac{\omega_4}{r^4}+\frac{\omega_5}{r^5}
+O\left(\frac{1}{r^6}\right),\label{w:inf1}
\end{eqnarray}
where $\omega_{\infty},~\omega_1,~M$ are three constants and the
coefficients $M_i$ and $\omega_i$ are expressed in terms of
$\omega_{\infty},~\omega_1,~M$, and $1+f'(R_0)$
\begin{eqnarray}
M_1&=&-\frac{(\omega_{\infty}^2-1)^2-\frac{\omega_1^2}{6}R_0}{2(1+f'(R_0))},\nn\\
M_2&=&\frac{2\omega_1\omega_{\infty}(1-\omega_{\infty}^{2})}{1+f'(R_0)},\nn\\
M_3&=&\frac{\omega_1^2 R_0
(1-2\omega_{\infty}^2)+8\omega_{\infty}^2(1-\omega_{\infty}^2)^2}{R_0
(1+f'(R_0))},\nn\\
M_4&=&\frac{-\omega_1^2(M+2\omega_1\omega_{\infty})
R_0+16\omega_1\omega_{\infty}
(1-\omega_{\infty}^{2})(2-3\omega_{\infty}^{2})}{2R_0(1+f'(R_0))},\nn\\
M_5&=&-\frac{1}{5R_0^2(1+f'(R_0))}\left[R_0^2\omega_1^4
-60MR_0\omega_1\omega_{\infty}
(-1+\omega_{\infty}^2)+24(-13+12\omega_{\infty}^2)(\omega_{\infty}-\omega_{\infty}^3)^2
\right.\nn\\
&&\left.-6R_0\omega_1^2(7-33\omega_{\infty}^2+26\omega_{\infty}^4)\right]
+\frac{\omega_1^2}{60(1+f'(R_0))^2}\left\{-R_0\omega_1^2+12(-1+\omega_{\infty}^2)^2\right\}\nn\\
\omega_2&=&\frac{6\omega_{\infty}(1-\omega_{\infty}^2)}{R_0},\nn\\
\omega_3&=&\frac{6\omega_1(1-\omega_{\infty}^2)}{R_0},\nn\\
\omega_4&=&\frac{-3(2M+\omega_1\omega_{\infty})\omega_1 R_0
+6\omega_{\infty}
(1-\omega_{\infty}^2)(7-3\omega_{\infty}^2)}{R_0^2},\nn\\
\omega_5&=&-\frac{3}{5R_0^2}\left[R_0\omega_1^3-96M\omega_{\infty}
(-1+\omega_{\infty}^2)
-6\omega_1(13-22\omega_{\infty}^2+9\omega_{\infty}^4)\right]\nn\\
&&-\frac{3\omega_1}{10R_0(1+f'(R_0))}\left\{R_0\omega_1^2-8(-1+\omega_{\infty}^2)^2\right\}.
\end{eqnarray}

\end{document}